\documentclass{article}
\usepackage{hyperref,amssymb,amsmath,mathrsfs,bm,graphicx}
\usepackage{epsfig,graphics}
\usepackage{color}
\begin{document}
\centerline{\large\bf AN EVOLUTION OF ADIABATIC MATTER:}
\centerline{\large\bf A CASE FOR THE QUASISTATIC REGIME}
\vspace*{0.245truein}
\centerline{\footnotesize{\large W. Barreto\footnote{Centro de F\'\i sica Fundamental,
Facultad de Ciencias, Universidad de Los Andes, M\'erida, Venezuela}}}
\baselineskip=12pt
\vspace*{0.21truein}
\date{\today}
\begin{abstract}
We establish the connection between the standard ADM 3+1 treatment of matter 
with its characteristic equivalent, in the context of spherical symmetry. 
The flux-conservative rendition of the fluid equations are obtained.
Considering adiabatic distributions of perfect fluid, we evolve the system
using the so-called post-quasi-static approximation in radiation coordinates.
We obtain an adiabatic matter evolution in the quasi-static regime or slow motion, which is not shear-free nor geodesic.

\vspace*{0.21truein}
\noindent Key words: Matter Evolution; Characteristic formulation for matter; Nonradiating evolution.
\end{abstract}
\section{Introduction}
{A unified approach to} the treatment of matter is desirable in numerical relativity, because it can be useful physically and geometrically. The standard way to consider matter in ADM 3+1 and characteristic formulations leads to flux--conservative equations \cite{nc00}, \cite{sfp01}. These
procedures are recognized as Eulerians \cite{font}.

{An old physical point of view to deal with matter \cite{b64}, combines Lagrangian and Eulerian observers, named Bondians after \cite{b09}; Bondian are comoving and local Minkowskian observers. {Observers in the ADM 3+1 treatment of matter are Bondian \cite{note}}. It is expected that this fact still holds true for the equivalent characteristic treatment of matter \cite{sfp01}.} 

{In the absence of spherical symmetry observers can be unambiguously introduced as comoving and local Minkowskian. However we do not have an unique way to choose the tetrad to go to the local Minkowskian spacetime, but
the way to go to the comoving with matter is straightforward \cite{melfoetal}.}

{Developing a general numerical solver of matter coupled to radiation,
within spherical symmetry, we {found} that the baryonic
matter behaves, near the regular center of symmetry, in a different way than a massless scalar field \cite{bcb09}.
Usually the scalar field is considered as a matter source \cite{bcgn02}. The massless scalar field can be interpreted as an anisotropic fluid \cite{bcb10}, but this is purely formal. At last the massless scalar field behaves as radiation does.
{Matter and radiation satisfy a central equation of state (CEoS)
in that asymptotic region which is conformally flat \cite{bcb09}}. Near $r=0$ we assume that the radial dependence of the geometrical and physical variables preserve the same dependence as the static variables near the coordinate origin. All this is valid for the characteristic treatment of a spherical distribution of  matter.}

{A general solver to evolve adiabatic matter is the first step towards the simulation of matter coupled to radiation. For that reason, 
we consider here the evolution of an adiabatic (nonradiating) fluid with a seminumerical procedure, which is a general enough method for the present purposes, as we shall see. 
Thus, we use the so-called post-quasi-static approximation (PQSA) \cite{hbds02,hjr80,hb11}. 
The PQSA  could be a tool for ADM 3+1 and characteristic formulations \cite{b09}.}

{We expect to develop alternative and general numerical hydrodynamic solvers which allow to handle selfgravitating and radiating matter evolutions. Namely, referential codes both in characteristic and ADM 3+1 formulations. {Our final goal is to apply these techniques to consider different transport mechanisms, dissipative effects as well as other degrees of freedom.} Coupling matter with radiation off spherical symmetry is relevant from the observational point of view, even for an isolated gravitational source.} 

{In section 2 we shall write the characteristic formulation of matter, that is, the field equations and the fluid dynamics equations, for an adiabatic spherical distribution of matter. We endeavor to maintain unity with the standard treatment of matter, showing the outcome using Bondian observers. Kinematical variables are introduced for analysis; an unambiguous matching conditions are considered in this section as well. {In section 3, for the sake of completeness, we shall present a r\'esum\'e of the PQSA;} in section 4 we shall use the PQSA to construct an adiabatic model. We shall conclude with some remarks in section 5. }
\section{Characteristic formulation of matter}
\subsection{Einstein equations}
Using radiation coordinates $(u,r,\theta,\phi)$, the line element adopts the form
\begin{equation}
ds^2=e^{2\beta}\left(\frac{V}{r}du^2+2dudr\right)-r^2(d\theta^2+\sin\theta^2d\phi^2) \label{LE},
\end{equation}
where $\beta$ and $V$ are functions of $u$ and $r$. 
It is useful to define the mass aspect function
\begin{equation}
m=\frac{1}{2}(r-e^{-2\beta}V), \label{mass2}
\end{equation}
which coincides with the Misner--Sharp mass \cite{ms64}.

Following Bondi \cite{b64}, we introduce purely local Minkowski coordinates $(t,x,y,z)$
\begin{equation}
dt=e^\beta\left[\sqrt{\frac{V}{r}} du+\sqrt{\frac{r}{V}}dr\right],\;\;dx=e^\beta \sqrt{\frac{r}{V}}dr,\;\;dy=rd\theta,\;\; dz=r\sin\theta d\phi.
\end{equation}
Next, when viewed by an observer moving relative to these coordinates
with local velocity $\omega=dx/dt$, the physical content of space
consists of {an isotropic fluid of energy--density $\rho$ and 
pressure $p$. The covariant energy--momentum tensor in the Minkowski coordinates is}
\begin{equation}
\left(
\begin{array}{cccc}
\rho & 0 & 0 & 0 \\
0 & p & 0 & 0 \\
0 & 0 & p & 0 \\
0 & 0 & 0 &  p 
\end{array}
\right).
\end{equation}
Thus, the matter velocity is
\begin{equation}
\frac{dr}{du}=\frac{V}{r}\frac{\omega}{1-\omega}, \label{drdu}
\end{equation}
and the Lorentz $\gamma$ function
\begin{equation}
\gamma^2=\frac{1}{1-\omega^2}.
\end{equation}
We define the conservative variables \cite{nc00}
\begin{equation}
\tau\equiv(\rho+p)\gamma^2-p=\frac{\rho+\omega^2 p}{1-\omega^2},
\end{equation}
\begin{equation}
S\equiv(\rho+p)\gamma^2\omega=\frac{(\rho+p)\omega}{1-\omega^2},
\end{equation}
and the flux variable
\begin{equation}
\kappa\equiv S\omega+p=\frac{p+\omega^2\rho}{1-\omega^2},
\end{equation}
which are related with the null (ingoing) flux--conservative variables \cite{nc00}, \cite{f06} by
\begin{equation}
\tau^-\equiv\tau-S=\frac{\rho-\omega p}{1+\omega}, \label{tau-}
\end{equation}
\begin{equation}
\kappa^-\equiv\kappa-S=\frac{p-\omega\rho}{1+\omega}, \label{kappa-}
\end{equation}
\begin{equation}
S^-\equiv\tau+\kappa-2S=\tau^-+\kappa^-. \label{S-}
\end{equation}
The nonzero components of the energy--momentum tensor are
\begin{equation}
T^u_u=\tau^-,\;\;T^r_r=-\kappa^-,\;\;T^u_r=\frac{r}{V}S^-,\;\;T^\theta_\theta=T^\phi_\phi=-p.
\label{Tes}
\end{equation}
The sufficient set of Einstein equations for the
variables $m$ and $\beta$ are given by the non--trivial component of the null momentum constraint (partial differentiation with respect to any coordinate is denoted by a comma)
\begin{equation}
m_{,u}=-4\pi r e^{2\beta}(r-2m)S. \label{nmc}
\end{equation}
the null polar slicing condition
\begin{equation}
\beta_{,r}=\frac{\;\;2\pi r^2 S^-}{r-2m }, \label{npsc}
\end{equation}
and the null Hamiltonian constraint
\begin{equation}
m_{,r}=4\pi r^2\tau^-. \label{nhc}
\end{equation}
Note that the null momentum constraint and the null Hamiltonian constraint are preserved in form with respect to the ADM 3+1.
This last equation can be easily integrated for any time. {Using  (\ref{nmc}) and (\ref{nhc}) we obtain} 
\begin{equation}
\frac{dm}{du}=-4\pi r^2 p \frac{dr}{du}, \label{dmdu}
\end{equation}
which is an energy equation (the power), showing clearly how the fluid pressure does work on a material sphere across its moving boundary. It can be easily shown that this last equation is exactly the first integral of the homogeneous equation of motion in the conservative form. 

The field equation  
$-8\pi T^\phi_\phi=-8\pi T^\theta_\theta=G^\phi_\phi=G^\theta_\theta$
which reads explicitly
\begin{eqnarray}
p&=&-\frac{e^{-2\beta}}{4\pi}\beta_{,ur} + \frac{1}{8\pi}\left(1-\frac{2m}{r}\right)
\left(2\beta_{,rr} + 4\beta_{,r}^2
 -\frac{\beta_{,r}}{r}\right) \nonumber \\ 
&&+\frac{1}{8\pi r}[3\beta_{,r}(1-2m_{,r}) - m_{,rr}], \label{e4}
\end{eqnarray}
can be written in many ways. To get some physical insight we can write it as a generalization of the well known Tolman--Oppenheimer--Volkoff (TOV) equation for hydrostatic support \cite{tov}, \cite{hjr80},
\begin{equation}
\kappa^-_{,r}+S^-\frac{(4\pi r^3\kappa^- + m)}{r(r-2m)}=\frac{2}{r}(p-\kappa^-)+e^{-2\beta}\left[\frac{rS^-}{r-2m}\right]_{,u}. \label{tov}
\end{equation}
or equivalently, as the inhomogeneous equation of motion for the fluid in conservative form, modulo Bianchi identities.
\subsection{Fluid dynamic equations}
{The basic equations of motion for the fluid can be derived from the
local conservation of the energy--momentum tensor, $T^{ab}_{;a}=0$, and the particle number, $(n\,u^a)_{;a}=0$, where ``;" is the (covariant) derivative operator compatible with $g_{ab}$. To these conservation laws we must adjoint an equation of state, $p=p(\rho_0,\epsilon)$, where $\rho_0$ is the rest mass energy--density and $\epsilon$ is the specific internal energy, which, further, must be consistent with the first law of thermodynamics. Thus, we are not considering any thermodynamics here (for a complete study see \cite{gourgo} and references therein).} 
When the equation of state is not a function of the number density, the time evolution of an ultra--relativistic perfect fluid is completely determined by the conservation of the stress--energy tensor.
The fluid equations of motion can be written in conservative form
\begin{equation}
\hat{\bf q}^-_{,u}+\frac{1}{r^2}(rV\hat{\bf f})_{,r}=\hat{\bf s}, \label{efcII}
\end{equation}
where the flux vector, $\hat{\bf f}$, and the source vector, $\hat{\bf s}$, are
\begin{equation}
\hat{\bf f}\equiv\left[\begin{array}{c}S \\ \kappa \end{array}\right],\;\;\;
\hat{\bf s}\equiv\left[\begin{array}{c}0 \\\varsigma\end{array}\right],
\end{equation}
the conservative variables vector, $\hat{\bf q}^-$, and primitive variables vector, $\hat{\bf w}$, are
\begin{equation}
\hat{\bf q}^-\equiv\left[\begin{array}{c}\;\;\;\tau^- \\-S^-\end{array}\right],\;\;\;
\hat{\bf w}\equiv\left[\begin{array}{c}p \\\omega\end{array}\right],
\end{equation}
\begin{equation}
\varsigma=\vartheta + \frac{2Vp}{r^2},
\end{equation}
and
\begin{equation}
\vartheta=e^{\beta}\left(\frac{V}{r}\right)^{1/2}\left[(S\omega-\tau)\left(8\pi r p + \frac{m}{r^2}\right)+p \frac{m}{r^2}\right].
\end{equation}
Observe that the equation (\ref{efcII}) entails (\ref{dmdu}) and (\ref{tov}).
{\subsection{Kinematical variables}
Kinematical variables such as the shear tensor and the four-acceleration will be useful for analysis in Section 4.  Being $g_{\mu\nu}$ the metric and $U^\mu$ the
four velocity, the shear tensor $\sigma_{\mu\nu}$ is given by 
\begin{equation}
\sigma_{\mu\nu}=U_{(\mu;\nu)}-U_{(\mu}A_{\nu)} - \frac{1}{3}\Theta P_{\mu\nu},
\end{equation}
where
\begin{equation}
\Theta=U^\mu_{;\mu}
\end{equation}
is the expansion,
\begin{equation}
P_{\mu\nu}=g_{\mu\nu}-U_\mu U_\nu
\end{equation}
is the projection tensor,
\begin{equation}
A_\nu=U^\mu U_{\nu;\mu}
\end{equation}
is the four--acceleration, satisfying the following conditions
$$\sigma_{\mu\nu}U^\mu=\sigma_{\mu\nu}g^{\mu\nu}=U^\mu A_\mu=0.$$ We can introduce the the scalars $\sigma$
and $A$ by means of
\begin{equation}
\sigma^2=\frac{1}{2}\sigma^{\mu\nu}\sigma_{\mu\nu}
\end{equation}
and
\begin{equation}
A^2=A^\mu A_{\mu}.
\end{equation}
For the present case, the metric (\ref{LE}) and the four velocity 
\begin{equation}
U^\mu= e^{-\beta}\left(\frac{r}{V}\right)^{1/2}\left(\frac{1-\omega}{1+\omega}\right)^{1/2}\delta^\mu_u+e^{-\beta}\left(\frac{V}{r}\right)^{1/2}\frac{\omega}{(1-\omega^2)^{1/2}}\delta^\mu_r
\end{equation}
lead us to
\begin{eqnarray}
\Theta&=&e^{-2\beta}\left[\left(\frac{1-\omega}{1+\omega}\right)^{1/2}(1-2m/r)^{3/2}\frac{m_{,u}}{r}
-\frac{\omega_{,u}(1-2m/r)^{-1/2}}{(1+\omega)(1-\omega^2)^{1/2}}\right]\nonumber\\
&+&\frac{(1-2m/r)^{1/2}\omega_{,r}}{(1-\omega^2)^{3/2}}+\frac{2\omega(1-2m/r)^{1/2}}{(1-\omega^2)^{1/2}}\left[\beta_{,r}+\frac{1}{r}\right]\nonumber\\
&+&\frac{\omega}{(1-\omega^2)^{1/2}(1-2m/r)^{1/2}}\left[\frac{m}{r^2}-\frac{m_{,r}}{r}\right],
\end{eqnarray}
\begin{equation}
A=A_re^{-\beta}\left(1+\frac{2V}{r}\frac{dr}{du}\right)^{1/2},
\end{equation}
where
\begin{eqnarray}
A_r&=&\frac{m_{,u}}{(1-2m/r)^{2}e^{2\beta}}\frac{(1-\omega)}{(1+\omega)}+\frac{m_{,r}}{(1+\omega)(1-\omega^2)(r-2m)}\nonumber\\
&-&\frac{m}{(1+\omega)r(r-2m)}-\frac{2\beta_{,r}}{(1+\omega)}-\frac{\omega\omega_{,r}}{(1+\omega)(1-\omega^2)}\nonumber\\
&-&\frac{\omega_{,u}}{(1-\omega^2)(1-2m/r)e^{2\beta}},
\end{eqnarray}
and
\begin{equation}
\sigma=\frac{\sqrt{3}}{r}\left[\frac{1}{3}\Theta r-\frac{\omega(1-2m/r)^{1/2}}{(1-\omega^2)^{1/2}}\right].
\end{equation}
With these variables, once the problem has been solved, we can observe if the evolution of the fluid is
geodesic or shear--free; we are not assuming these features.
\subsection{Junction conditions}
{We have to match the interior (dynamic) solution with the exterior solution, which is static by virtue of the Birkoff theorem. 
Thus, outside of the fluid distribution, the spacetime is that of Schwarzschild
\begin{equation}
ds^{2}=\left(1-\frac{2M}{r}\right)du^2+2dudr-r^2(d\theta^2+\sin\theta d\phi^2),
\end{equation}
where $M$ is the total mass.
The boundary conditions at some moving radius are that of DarmoisÐ-Lichnerowicz \cite{darmois-lichne}. These are equivalent to the continuity of the first and second fundamental forms \cite{hj83}. 
In order to match smoothly the two metrics at the surface $r = r_\Sigma(u)$,
the functions $\beta$ and $m$, are continuous across the boundary of the sphere
\begin{equation}
\beta_\Sigma=0,\;\;m_\Sigma=M, \label{fff}
\end{equation}
where the subscript $\Sigma$ indicates the boundary of the surface distribution.
The continuity of the second fundamental form leads us to
\begin{equation}
\left[-\beta_{,u}e^{2\beta}+\left(1-\frac{2m}{r}\right)\beta_{,r}-\frac{m_{,r}}{2r}\right]_\Sigma=0
\end{equation}
which is equivalent to
\begin{equation}
p_\Sigma=0,\label{pa} 
\end{equation}
and expresses the continuity of the pressure at the surface of the matter distribution.
{In case of dissipation the radial pressure is discontinuous \cite{s85}, \cite{bhj89}, being proportional to the heat flow and/or to the shear
viscosity at the boundary.} 

Up to this point the treatment of matter is completely general, in the context of
spherical symmetry, isotropic and adiabatic fluids. However, in the next section we {show how the system  can be integrated semi--numerically using the PQSA.}
{\section{A seminumeric method}
To introduce the PQSA procedure some general considerations will be necessary.
\subsection{Equilibrium and departures from equilibrium}
The simplest situation, when dealing with self--gravitating spheres, is that of equilibrium (static case). Next, we have the quasistatic regime. By this we mean that the sphere changes slowly, on a time scale that is very long compared to the typical time in which the sphere reacts to a slight perturbation of hydrostatic equilibrium. This typical time scale is called the hydrostatic time scale.
In the quasistatic regime the system remains in (or very close to) equilibrium. 
{This assumption is very sensible because the hydrostatic time scale is very small for many phases of the life of the star. It is of the order of 27 min
for the Sun, 4.5 s for a white dwarf, and $10^{-4}$ s for a neutron star of one solar mass and 10 km radius. It is well known that all the stellar configurations mentioned above generally change on time scales that are very long compared to their respective hydrostatic time scales \cite{hbds02}.}
However, during their evolution, self--gravitating objects may pass through phases of intense dynamical activity, with time scales of the order of magnitude of (or even smaller than) the hydrostatic time scale, and for which the quasistatic approximation is clearly not reliable{, as in the collapse of very massive stars and the quick collapse phase which precedes the formation of neutron stars \cite{i63}, \cite{mb90}.}

In these cases it is mandatory to consider departures from equilibrium {(see Ref. \cite{hbds02} and references therein for details about the translation of the assumption into conditions on the radial local velocity and the metric functions).}
\subsection{The effective variables and the Postquasistatic method}
Let us define the flux--conservative variables (\ref{tau-}) and (\ref{kappa-}) as the effective energy and pressure, respectively. Observe that in the static case they coincide with the energy density and the pressure. Thus, in the quasistatic situation (and obviously in the static too), effective and physical variables share the same radial dependence, {satisfying the same hydrostatic support equation}. Next, feeding back Eqs. (\ref{tau-}) and (\ref{kappa-}) into Eqs. (\ref{npsc}) and (\ref{nhc}), these two equations may be formally integrated, to obtain
\begin{equation}
\beta=2\pi\int^r_{r_\Sigma} \frac{(\tau^-+\kappa^-)}{r-2m } r^2 dr
\end{equation}
and
\begin{equation}
m=4\pi\int^r_0  r^2\tau^- dr.
\end{equation}
From here it is obvious that for a given radial dependence of the effective variables the radial dependence of the metric functions become completely determined. With this last comment in mind, we shall define the postquasistatic regime as that corresponding to a system out of equilibrium (or quasiequilibrium) but whose effective variables share the same radial dependence as the corresponding physical variables in the state of equilibrium (or quasiequilibrium). Alternatively, it may be said that the system in the postquasistatic regime is characterized by metric functions whose radial dependence is the same as the metric functions corresponding to the static (quasistatic) regime. The rationale behind this definition is not difficult to grasp: we look for a regime which, although out of equilibrium, represents the closest possible situation to a quasistatic evolution.
\subsection{The algorithm}
{Let us now outline the approach for the adiabatic case:
\begin{enumerate}
\item  Take an interior solution to Einstein equations, representing a fluid
distribution of matter in equilibrium, with a given $\rho_{st}=\rho(r)$ and $p_{st}=p(r)$.
\item  Assume that the $r$ dependence of $\kappa^-$ and $\tau^-$ is the
same as that of $p_{st}$ and $\rho_{st}$, respectively.
\item  Using equations (\ref{npsc}) and (\ref{nhc}), with the $r$ dependence of
$\kappa^-$ and $\tau^-$, one gets $\beta$ and $m$ up to some functions of
$u$, which will be specified below.
\item  For these functions of $u$ one has three ordinary differential equations
(hereafter referred to as surface equations), namely, equations (\ref{drdu}), (\ref{dmdu}) and (\ref{tov}) evaluated on $r=r_{\Sigma}$.
\item Once the system of surface equations is determined, it may be integrated for
any particular initial data set.
\item  Feeding back the result of integration in the expressions for $\beta$ and
$m$, these two functions are completely determined.
\item  With the input from the point 6 above, and using field equations,
 all physical variables may be found for any piece of
matter distribution.
\end{enumerate}}
{We shall see in the next section that the PQSA is general enough in the present context, due to the physical features displayed.}}
\section{An adiabatic evolution}
{Assuming that the conservative variable $\tau^-$ is an exclusive function of time we can solve the system  using the PQSA.}
We build a Schwarzschild--like model which corresponds to an incompressible isotropic fluid as the static ``seed'' interior solution. Thus, the {so-called} effective energy density and the effective pressure, respectively, are
\begin{equation}
\tau^-=f(u)=\frac{3m_\Sigma}{4\pi r_\Sigma^3}\label{eed}
\end{equation}
and
\begin{equation}
\kappa^-={\tau^-}\left\{\frac{(1-3\omega_\Sigma)\xi-(1-\omega_\Sigma)\xi_\Sigma}
{3(1-\omega_\Sigma)\xi_\Sigma-(1-3\omega_\Sigma)\xi}\right\},\label{erp}
\end{equation}
where
$$\xi=\left[1-\frac{2m_\Sigma}{r_\Sigma}\left(\frac{r}{r_\Sigma}\right)^2\right]^{1/2}.$$
{Observe that in the static case the effective variables (\ref{eed}) and (\ref{erp}) (the ingoing flux-conservative variables (\ref{tau-}) and (\ref{kappa-})) coincide exactly with the energy density and the pressure, respectively.}
With these effective variables and integrating (\ref{npsc}) and (\ref{nhc}) we obtain
\begin{equation}
\beta=\frac{1}{2}\ln\left\{(1-\omega_\Sigma)\left[\left(\frac{3}{2}\frac{\xi_\Sigma}{\xi}-\frac{1}{2}\right)\right]+\omega_\Sigma\right\}
\end{equation}
and
\begin{equation}
m={m_\Sigma}\left(\frac{r}{r_\Sigma}\right)^3.
\end{equation}
The system of equations is reduced to a set at the surface, these are, Eqs. (\ref{drdu}), (\ref{dmdu}) and (\ref{tov}), and all the variables can be written in terms of $r_\Sigma$, $m_\Sigma$ and $\omega_\Sigma$, for which we need initial values to proceed with numerical integrations. By virtue of Eq. (\ref{dmdu}) and the boundary condition (\ref{pa}), the total mass remains constant for the adiabatic case.
In geometrized units the following initial values represent a compact/noncompact distributions of matter
with an initial contracting velocity:
$$r_\Sigma(0)=5M; 50M,\;\;\;\omega_\Sigma(0)=-10^{-5}.$$
Using a fourth order Runge--Kutta integrator, with a time step $\Delta u=10^{-3}$, we solve the system of two ordinary differential equations obtained from (\ref{drdu}) and (\ref{tov}) evaluated at the surface. To determine at any piece of the material $\rho$, $p$ and $\omega$, including $r=0$, we use Eqs. (\ref{nmc}), (\ref{npsc}) and (\ref{nhc}), but checking carefully that (\ref{e4}) holds. For this reason we define the simple error $\epsilon=\omega\Delta p$, where  $\Delta p$ is the difference between the pressure calculated using {(\ref{e4})} and the pressure using {(\ref{nmc}), (\ref{npsc}) and (\ref{nhc})}.
\begin{figure}[!ht]
\includegraphics[width=5.in,height=6.in,angle=0]{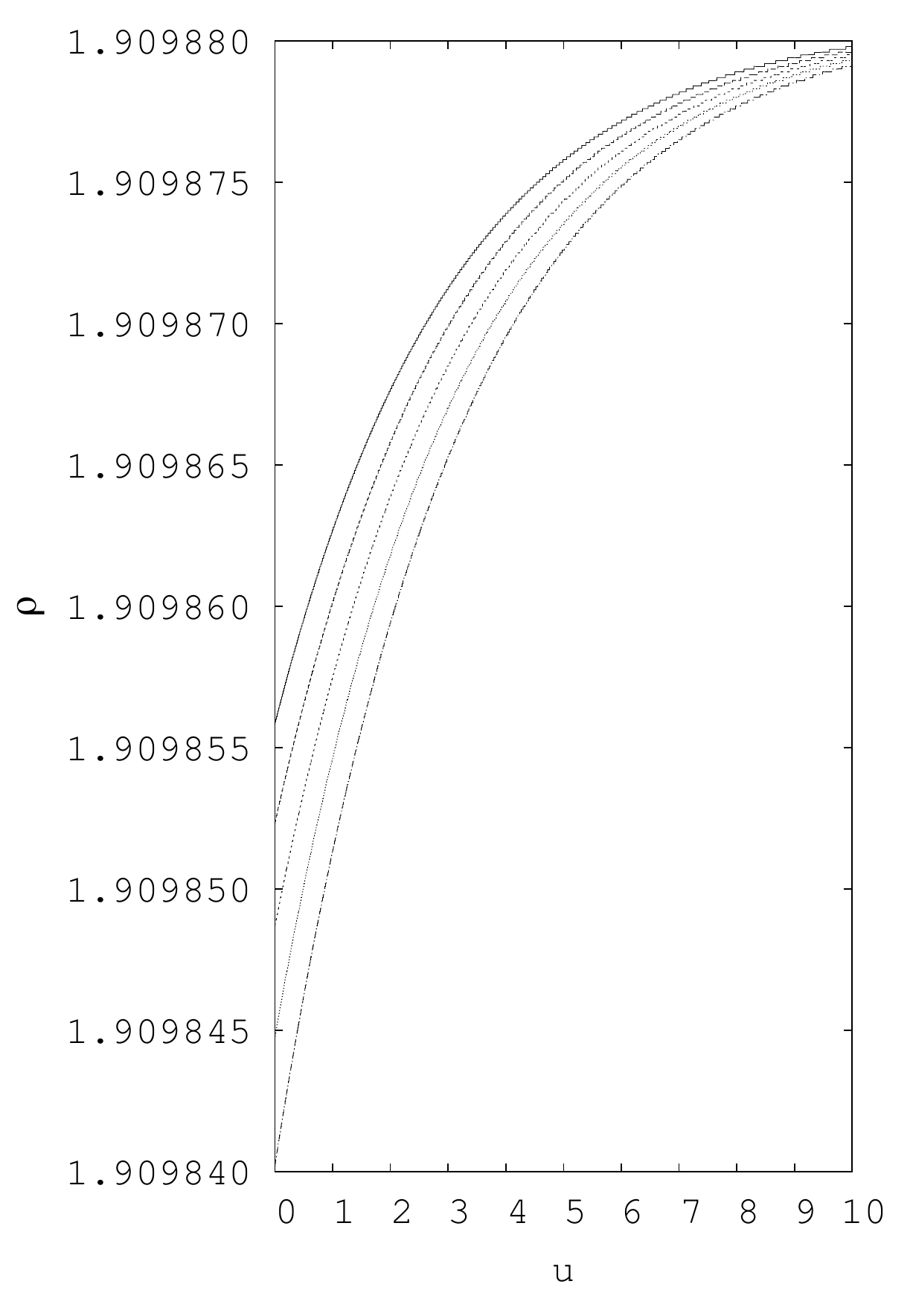}
\caption{Energy density $\rho$ (multiplied by $10^3$) as a function of the Bondi time for different regions: $r/r_\Sigma=0.2$ (continuos line); $r/r_\Sigma=0.4$ (dashed line); $r/r_\Sigma=0.6$ (small dashed); $r/r_\Sigma=0.8$ (dotted line); $r/r_\Sigma=1.0$ (dash--dotted line). The situation is
compact ($r_\Sigma(0)=5M$).}
\label{fig:density}
\end{figure}
\begin{figure}[!ht]
\includegraphics[width=5.in,height=6.in,angle=0]{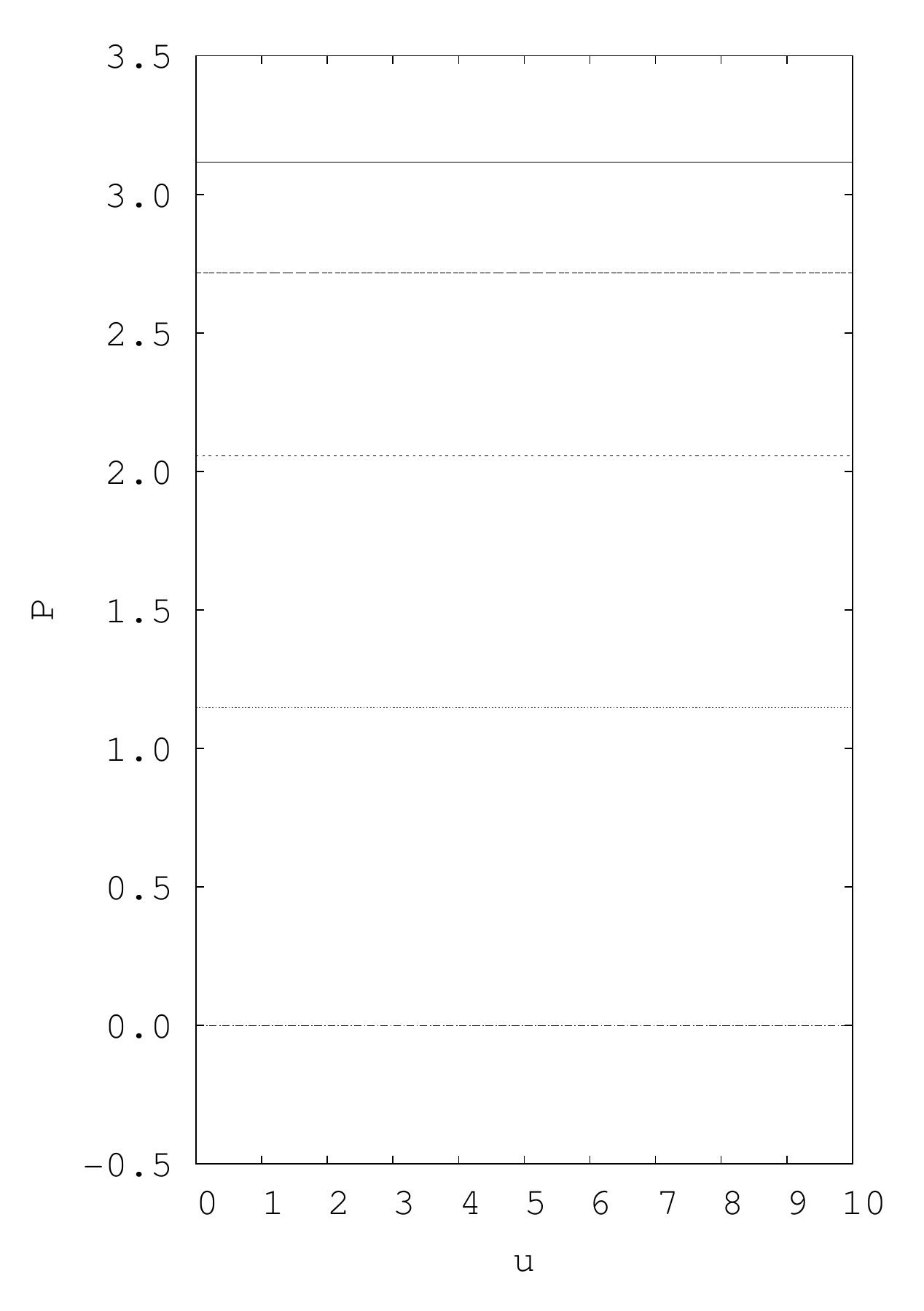}
\caption{Pressure $p$ (multiplied by $10^4$) as a function of the Bondi time for different regions: $r/r_\Sigma=0.2$ (continuos line); $r/r_\Sigma=0.4$ (dashed line); $r/r_\Sigma=0.6$ (small dashed); $r/r_\Sigma=0.8$ (dotted line); $r/r_\Sigma=1.0$ (dash--dotted line). The situation is
compact ($r_\Sigma(0)=5M$).}
\label{fig:pressure}
\end{figure}
\begin{figure}[!ht]
\includegraphics[width=5.in,height=6.in,angle=0]{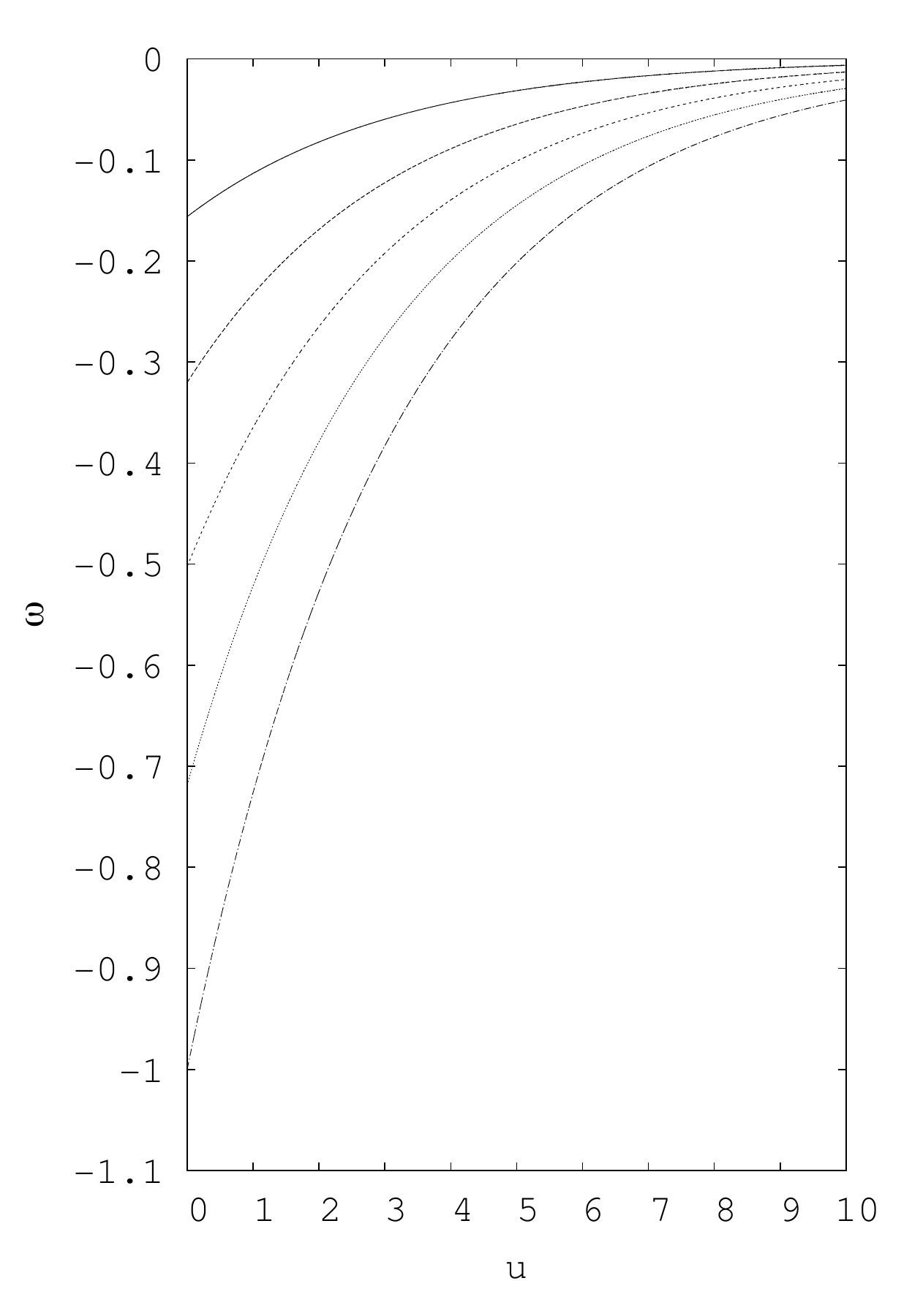}
\caption{Local velocity $\omega$ (multiplied by $10^5$) as a function of the Bondi time for different regions: $r/r_\Sigma=0.2$ (continuos line); $r/r_\Sigma=0.4$ (dashed line); $r/r_\Sigma=0.6$ (small dashed); $r/r_\Sigma=0.8$ (dotted line); $r/r_\Sigma=1.0$ (dash--dotted line). The situation is
compact ($r_\Sigma(0)=5M$).}
\label{fig:velocity}
\end{figure}
\begin{figure}[!ht]
\includegraphics[width=5.in,height=6.in,angle=0]{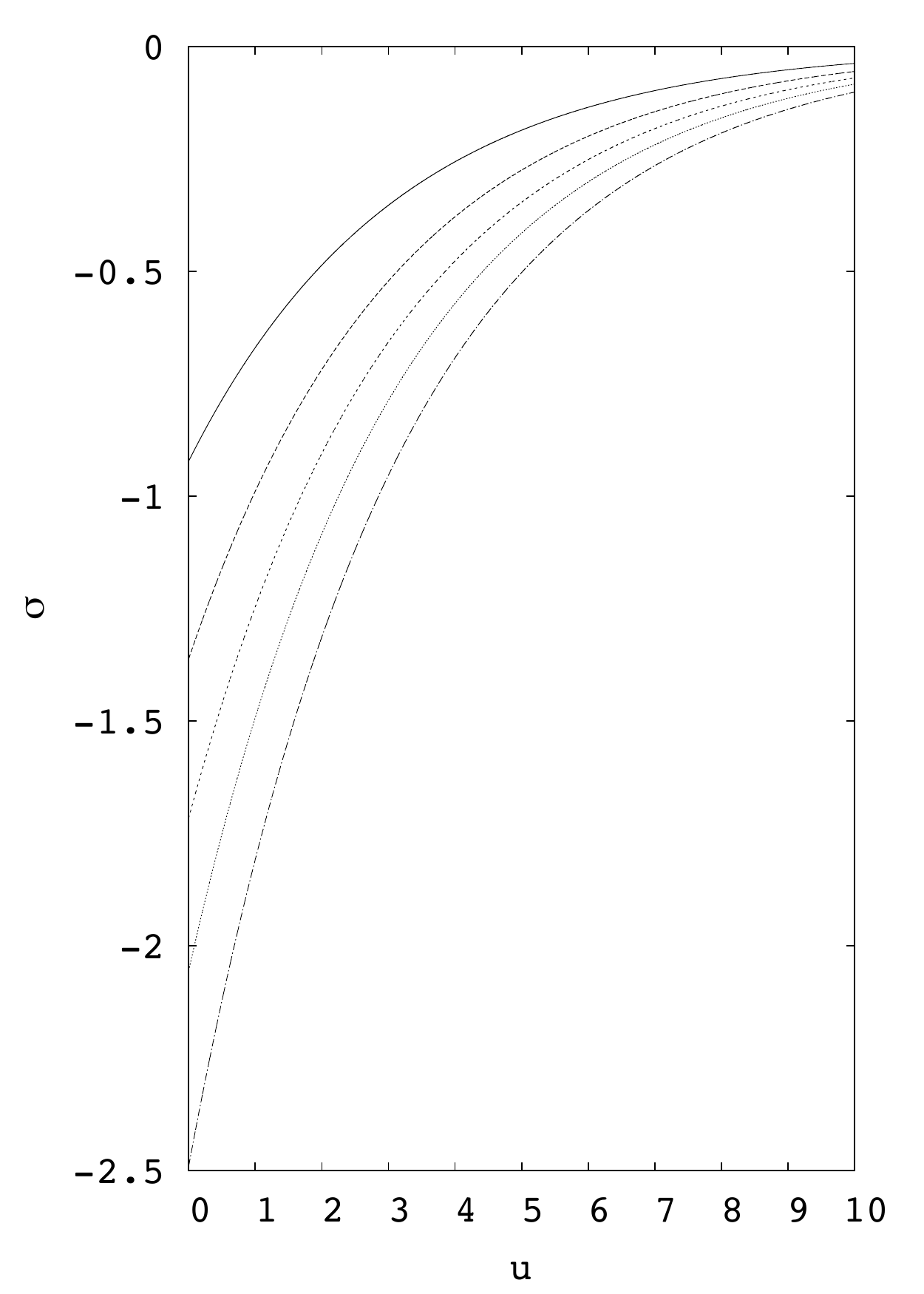}
\caption{Shear scalar $\sigma$ (multiplied by $10^6$) as a function of the Bondi time for different regions: $r/r_\Sigma=0.2$ (continuos line); $r/r_\Sigma=0.4$ (dashed line); $r/r_\Sigma=0.6$ (small dashed); $r/r_\Sigma=0.8$ (dotted line); $r/r_\Sigma=1.0$ (dash--dotted line). The situation is
compact ($r_\Sigma(0)=5M$).}
\label{fig:shear}
\end{figure}

\begin{figure}[!ht]
\includegraphics[width=5.in,height=6.in,angle=0]{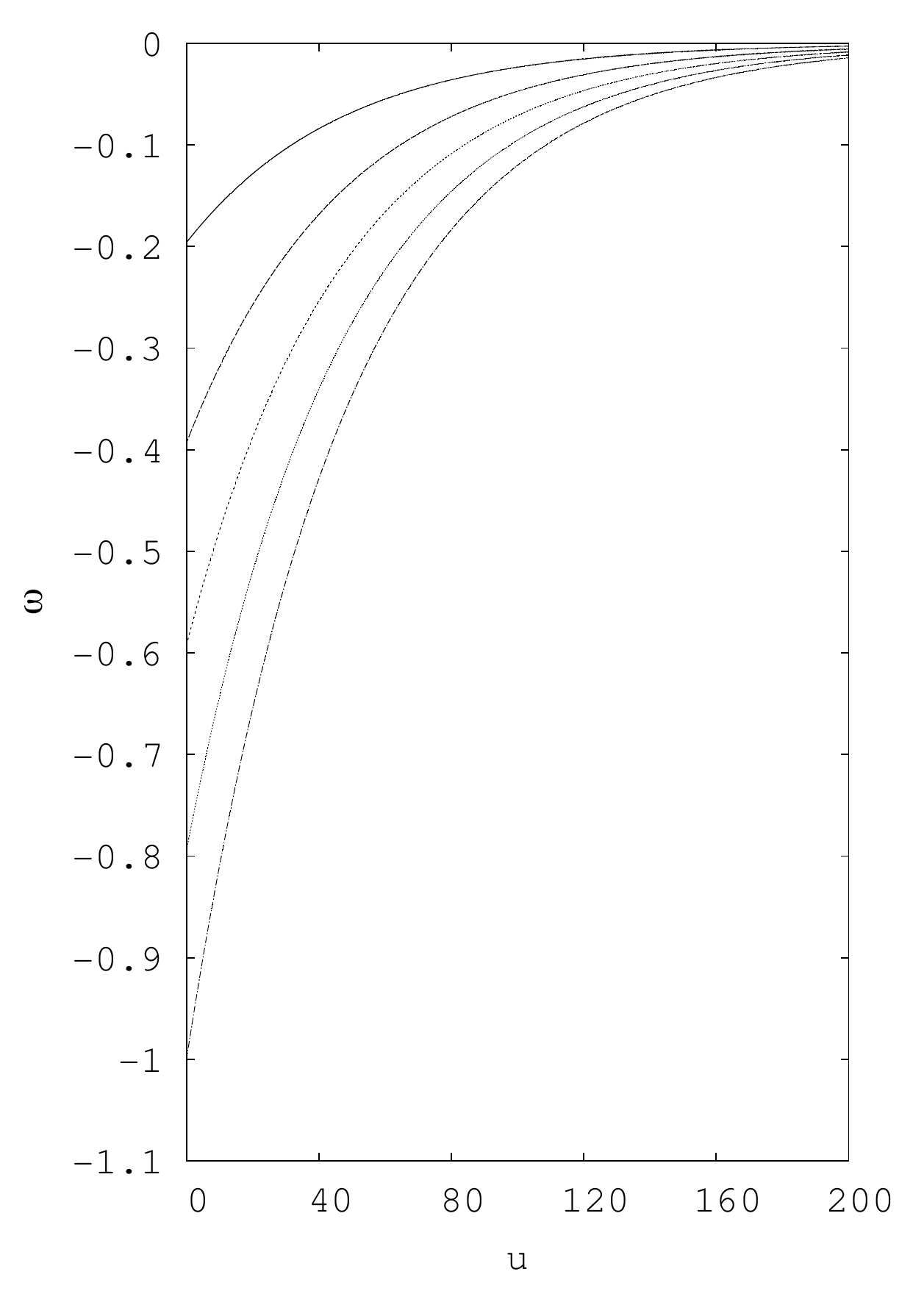}
\caption{Local velocity $\omega$ (multiplied by $10^5$) as a function of the Bondi time for different regions: $r/r_\Sigma=0.2$ (continuos line); $r/r_\Sigma=0.4$ (dashed line); $r/r_\Sigma=0.6$ (small dashed); $r/r_\Sigma=0.8$ (dotted line); $r/r_\Sigma=1.0$ (dash--dotted line). The situation is
less compact ($r_\Sigma(0)=50M$).}
\label{fig:velocity_less_compact}
\end{figure}
\begin{figure}[!ht]
\includegraphics[width=5.in,height=6.in,angle=0]{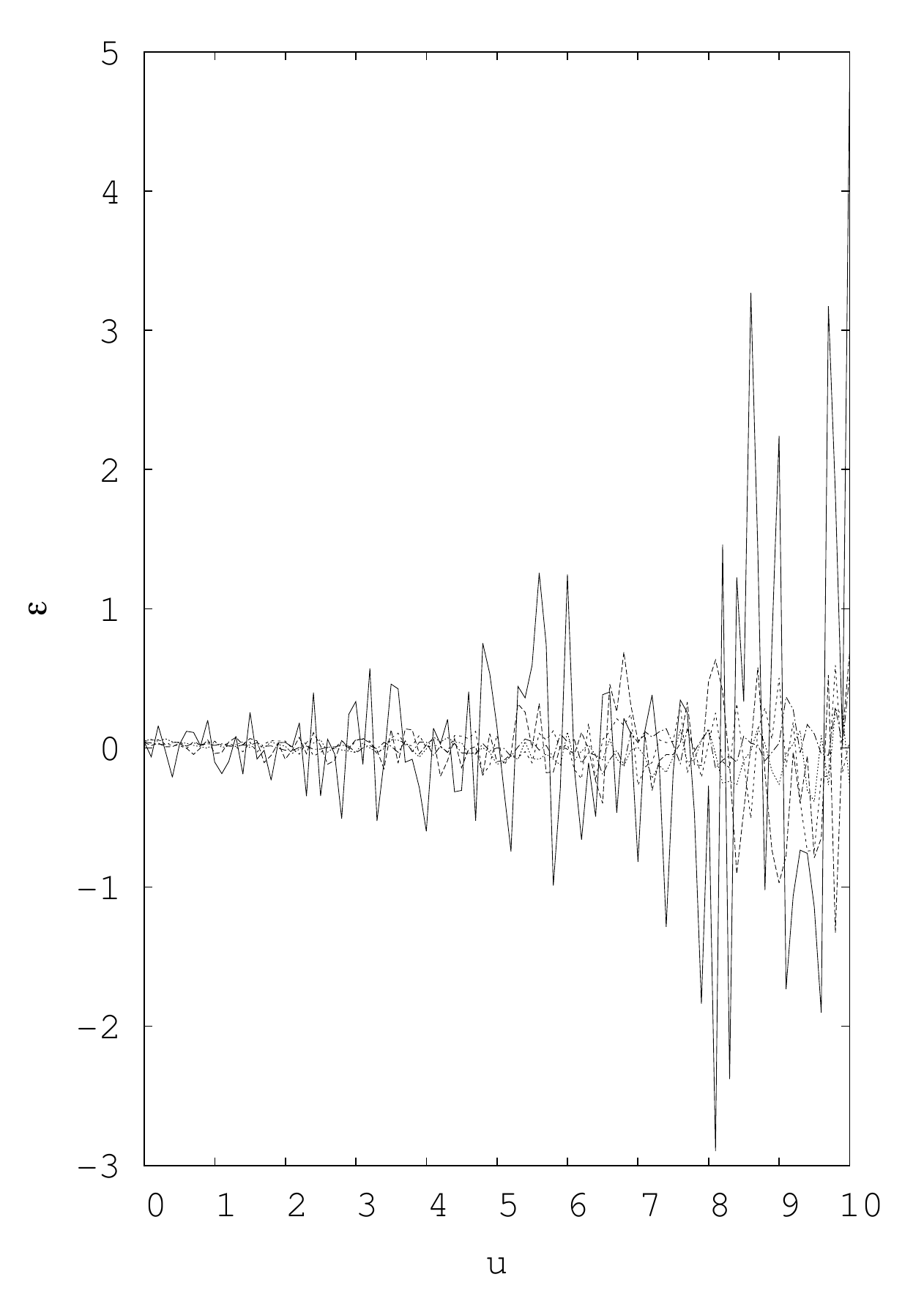}
\caption{Error (multiplied by $10^{12}$) about ${\cal O}(\Delta u^4)$ in calculating $\omega \Delta p$ as a function of the Bondi time for different regions: $r/r_\Sigma=0.2$ (continuos line); $r/r_\Sigma=0.4$ (dashed line); $r/r_\Sigma=0.6$ (small dashed); $r/r_\Sigma=0.8$ (dotted line); $r/r_\Sigma=1.0$ (dash--dotted line). }
\label{fig:error}
\end{figure}
\begin{figure}[!ht]
\includegraphics[width=5.in,height=6.in,angle=0]{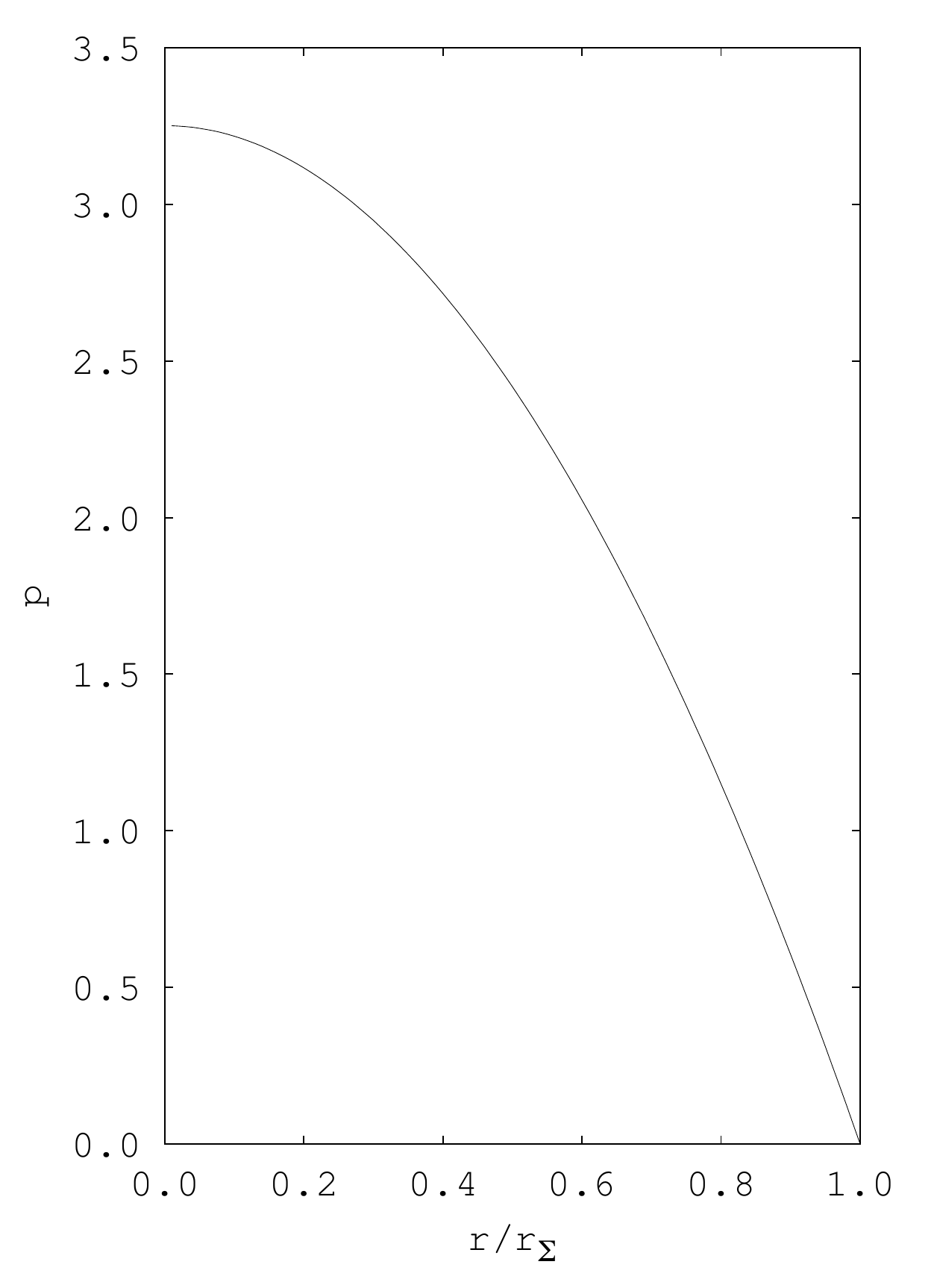}
\caption{Pressure $p$ (multiplied by $10^4$) as a function of $r/r_\Sigma$ for any time.  The situation is compact ($r_\Sigma(0)=5M$).}
\label{fig:p}
\end{figure}
\begin{figure}[!ht]
\includegraphics[width=5.in,height=6.in,angle=0]{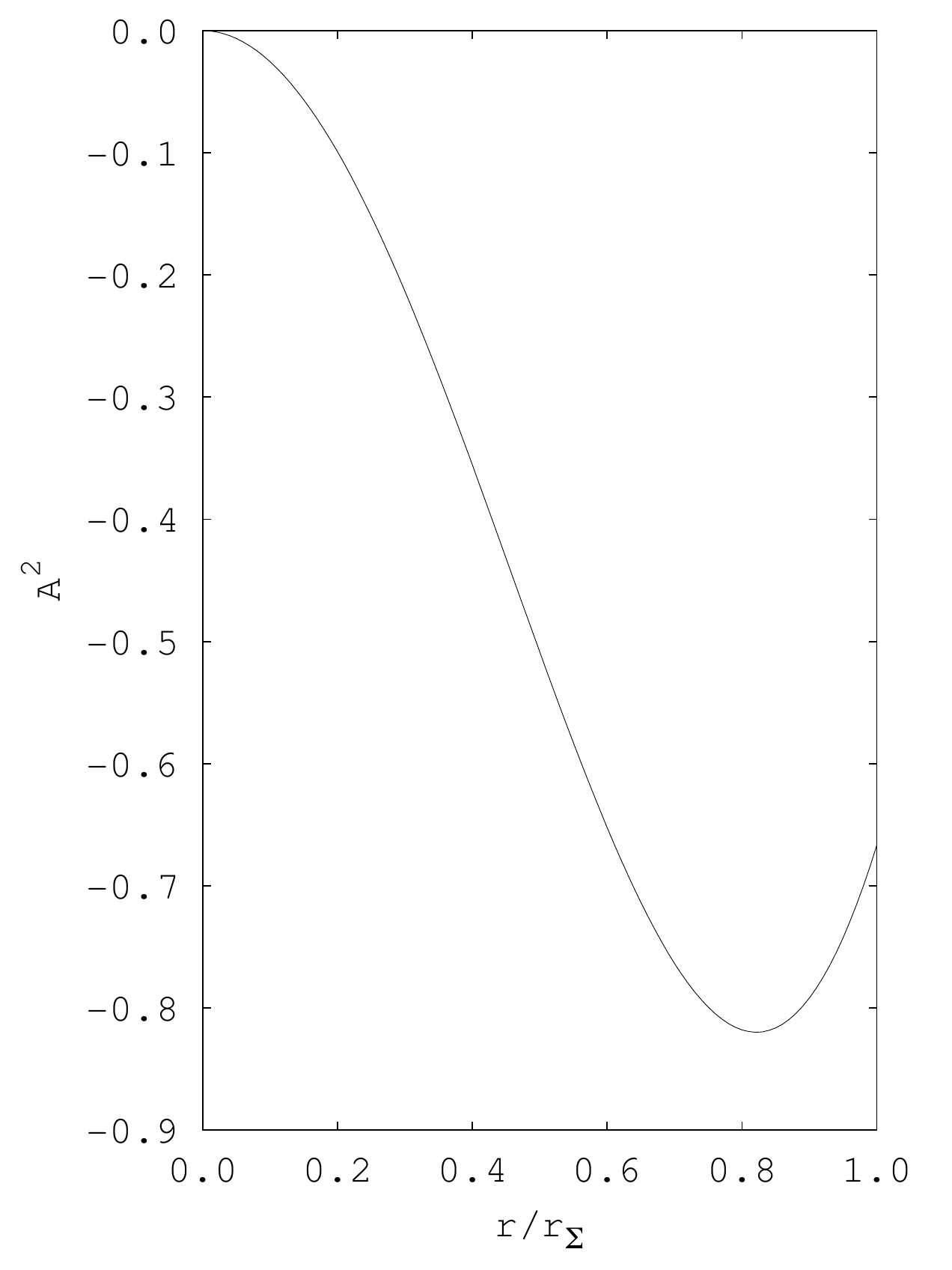}
\caption{Four acceleration scalar $A^2$ (multiplied by $10^3$) as a function of $r/r_\Sigma$ for any time.  The situation is compact ($r_\Sigma(0)=5M$).}
\label{fig:A}
\end{figure}

Figures 1--4 show the profiles for $\rho$, $p$, $\omega$ {and $\sigma$}, as a function of time for different pieces of comoving shells in a compact situation ({$r_\Sigma(0)=5M$}). Figure 5 shows the local velocity as a function of time for different pieces of comoving shells in a lesser compact situation ({$r_\Sigma(0)=50M$}).  Figure 6 shows the controlled error $\epsilon$ for the quasistatic evolution. {Figure 7 and 8 display the pressure and the four acceleration as a function of $r/r_\Sigma$ for any time.}

{From results, the energy density goes to a constant value, consistent with the incompressible static ``seed''; that is, with evolution the system recovers the static situation (asymptotically). Although the gradient of pressure is noticeable, the pressure at each shell $r/r_\Sigma$ is apparently constant. {We observe $\rho > p$ and that the fluid is not shear--free at any piece of material.} 
Other feature of the quasi-static regime exhibited in the present evolution is the following. Less compact is the system, the static equilibrium is recovered later (although the system seems to recover static equilibrium in an infinite Bondi time).

A numerical artifact in our calculations is a noise introduced when the local velocity is too small. This was not filtered but monitored under control. In all cases the evolution is necessarily slow (quasistatic), otherwise the field equation (\ref{e4}) is not satisfied. 

The profiles of $p$ and $A$ are better displayed as a function of $r/r_\Sigma$ for any time. They apparently do not change in time at each $r/r_\Sigma$, but clearly depend on $r/r_\Sigma$. The system neither evolves geodesically.}

{The initial state of the system is determined in practice by the initial conditions to integrate Eqs. (\ref{drdu}) and (\ref{tov}) evaluated on $r=r_\Sigma$.
Thus, the initial radial local velocity, $\omega_\Sigma(0)$, can be considered
as a perturbative agent. From this point of view, the energy density $\rho$ and the pressure $p$ are also perturbed from equilibrium. {One can think that the system must oscillate instead to be ``critically damped''. This is not the case, as far we see, because the radial dependence is preserved in the PQSA (see section 3). We are seeing only how the system is leaving equilibrium.} 

{Without the pretension of modeling specific astrophysical scenarios, we have presented one example, in the simplest (adiabatic) case.
In this model the profile of the shear tensor and the four acceleration clearly illustrate the ``dynamics'' of the model, tending to zero in the static limit. The fact that the shear tend to vanish in the quasistatic regime (for this specific model) further brings out their relevance in the treatment of situations off equilibrium. On the other hand, the velocity profiles show the relativistic gravitational effects in the quasistatic regime.}

\section{Concluding remarks}
We considered in this paper the characteristic treatment of adiabatic (non radiating) matter,
using Bondian observers. This leads us straightforwardly to an ADM 3+1--like and flux-conservative rendition of equations, which is appropriate for developing general an alternative numerical solvers. However, we solved seminumerically  the system of equations using the so-called PQSA, which is a good approach for
the evolution of adiabatic matter. 
  
{An adiabatic evolution of an ``incompressible'' fluid can be slow, in the context of spherical symmetry using the radiation coordinates. Such an evolution is not shear--free nor geodesic.}

{A general solver for evolving adiabatic matter is the first step towards the simulation of matter coupled to radiation.  At last we expect to make a detailed account of the fluid dynamics in the gravitational collapse considering inner dissipative transport mechanisms as diffusion and free streaming, as well as viscosity, anisotropy and electric charge. Work in this direction is in progress.}

\section*{Acknowledgments} 
The author wishes to thank Luis Herrera, Carlos Peralta and Luis Rosales. {To the Departamento de F\'\i sica Te\'orica e Historia de la Ciencia, Universidad del Pa\'\i s Vasco, for hospitality, and also the Intercambio Cient\'\i fico Program, Universidad de Los Andes, for financial support.

{\bibitem{note} Authors assure (Romero et al.: Ap. J. {\bf 462}, 839 (1996)) that the Schwarzschild type coordinates allow a simple extension of the Eulerian Newtonian hydrodynamics to the Einsteinian one. It refers to Bondi's way to deal with matter in General Relativity
(see Ref. \cite{b64}). That is also true in radiation coordinates.}
{\bibitem{melfoetal} Herrera, L., Melfo, A. N\'u\~nez, L. Pati\~no, A.: Ap. J. {\bf 421}, 677 (1994)}

\bibitem{bcb09} Barreto, W., Castillo, L., Barrios, E.: Phys. Rev. D {\bf 80}, 084007 (2009)
\bibitem{bcgn02} Brady, P. R., Choptuik, M. W., Gundlach, C.,  Neilsen, D. W.: Class. Quant. Grav. {\bf 19} 6359 (2002)
\bibitem{bcb10} Barreto, W., Castillo, L., Barrios, E.: General Relativity and Gravitation, {\bf 42} 1845 (2010)
\bibitem{hbds02} Herrera, L., Barreto, W., Di Prisco, A., Santos, N. O.: Phys. Rev. D {\bf 65}, 104004 (2002)
\bibitem{hjr80} Herrera, L., Jim\'enez, J.,  Ruggeri, G. J.: Phys. Rev. D {\bf 22}, 2305 (1980)
\bibitem{hb11} Herrera, L., Barreto, W.: Int. J. Mod. Phys. D {\bf 20}, 1265 (2011)
\bibitem{ms64} Misner, C., Sharp, D.: Phys. Rev. {\bf 136}, B571 (1964)
\bibitem{f06} Frittelli, S.: Phys. Rev. D, {\bf 73}, 124001 (2006)
\bibitem{tov} Tolman, R.: Phys. Rev. {\bf 55}, 364 (1939); Oppenheimer, J.,  Volkoff, G.: Phys. Rev. {\bf 55}, 374 (1939)
\bibitem{gourgo} Gourgoulhon, E.: Astron. Astrophys. {\bf 252}, 651 (1991)
\bibitem{darmois-lichne} Darmois, G.: Memorial des Sciences Mathematiques (GauthierÐVillars, Paris, 1927), Fasc. 25; Lichnerowicz, A.: Theories Relativistes de la Gravitation et de l'Electromagnetisme (Masson, Paris, 1955)
\bibitem{hj83} Herrera, L., Jim\'enez, J.: Phys. Rev. D {\bf 28}, 2987 (1983)
\bibitem{s85} Santos, N. O.: Mon. Not. R. Astron. Soc. {\bf 216}, 403 (1985)
\bibitem{bhj89} Herrera, L., Barreto, W, Jim\'enez, J.: Can. J. Phys. Can. J. Phys. {\bf 67}, 885 (1989)
\bibitem{i63} Iben, I.: Astrophys. J. {\bf 138}, 1090 (1963)
\bibitem{mb90} Myra, E., Burrows, A.: Astrophys. J. {\bf364}, 222 (1990)
\end{document}